\documentclass[conference]{IEEEtran}
\IEEEoverridecommandlockouts
\usepackage{cite}
\usepackage{amsmath,amssymb,amsfonts}
\usepackage{algorithmic}
\usepackage{graphicx}
\usepackage{textcomp}
\usepackage{xcolor}
\usepackage{booktabs}
\def\BibTeX{{\rm B\kern-.05em{\sc i\kern-.025em b}\kern-.08em
    T\kern-.1667em\lower.7ex\hbox{E}\kern-.125emX}}
\begin{document}

\title{On in-silico estimation of left ventricular end-diastolic pressure from cardiac strains\\

}

\author{\IEEEauthorblockN{Emilio A. Mendiola}
\IEEEauthorblockA{\textit{Department of Biomedical Engineering} \\
\textit{Texas A\&M University}\\
College Station, TX, USA\\
Email: emilio.mendiola@tamu.edu}
\and
\IEEEauthorblockN{Raza Rana Mehdi}
\IEEEauthorblockA{\textit{Department of Biomedical Engineering} \\
\textit{Texas A\&M University}\\
College Station, TX, USA\\
Email: razamehdi@tamu.edu}
\and
\IEEEauthorblockN{Dipan J. Shah}
\IEEEauthorblockA{\textit{DeBakey Heart and Vascular Center} \\
\textit{Houston Methodist}\\
Houston, TX, USA}
Email: adarwish@houstonmethodist.org
\and
\IEEEauthorblockN{Reza Avazmohammadi}
\IEEEauthorblockA{\textit{Department of Biomedical Engineering} \\
\textit{Texas A\&M University}\\
College Station, TX, USA \\
Email: rezaavaz@tamu.edu}

}

\maketitle

\begin{abstract}
Left ventricular diastolic dysfunction (LVDD) is a group of diseases that adversely affect the passive phase of the cardiac cycle and can lead to heart failure. While left ventricular end-diastolic pressure (LVEDP) is a valuable prognostic measure in LVDD patients, traditional invasive methods of measuring LVEDP present risks and limitations, highlighting the need for alternative approaches. This paper investigates the possibility of measuring LVEDP non-invasively using inverse in-silico modeling. We propose the adoption of patient-specific cardiac modeling and simulation to estimate LVEDP and myocardial stiffness from cardiac strains. We have developed a high-fidelity patient-specific computational model of the left ventricle. Through an inverse modeling approach, myocardial stiffness and LVEDP were accurately estimated from cardiac strains that can be acquired from in vivo imaging, indicating the feasibility of computational modeling to augment current approaches in the measurement of ventricular pressure. Integration of such computational platforms into clinical practice holds promise for early detection and comprehensive assessment of LVDD with reduced risk for patients.
\end{abstract}


\section{Introduction}

Left ventricular diastolic dysfunction (LVDD) involves an impairment in the ability of the left ventricle to relax and fill adequately during the diastolic phase of the cardiac cycle. Relaxation impairment is often coincident with increased myocardial stiffness and can lead to chronic heart failure \cite{Jeong2016}. Elevated LVEDP serves as a key indicator of the extent of diastolic dysfunction, reflecting reduced ventricular compliance or impaired relaxation \cite{Lalande2008,Tanmay2023LV}. Clinically, assessing LVEDP is crucial for diagnosing and grading the severity of diastolic dysfunction \cite{Jeong2016}. However, current invasive methods of measuring LVEDP are costly and incur risks and discomfort in patients. 

In the normal heart, the myocardial architecture and composition generate optimal passive stretching in diastole. Alterations in the mechanical behavior of the ventricle leading to impaired filling coincide with alterations in the kinematic behavior of the myocardium. Both passive and active kinematics of the LV are innately linked to the myocardium structural and mechanical properties, which are altered in the presence of structural diseases such as LVDD \cite{Mendiola2023,Mendiola2024}. Previous simulation-based investigation has indicated the feasibility of predicting architectural metrics from regional strain data \cite{Usman2023}. Given such intrinsic connections between myocardium behavior and regional deformation, we hypothesize a patient-specific computational model of the LV is capable of predicting the ventricular load and tissue stiffness from cardiac strains that are readily available from standardized imaging sequences.

In this work, we propose a patient-specific modeling approach to be used as a non-invasive alternative to estimate myocardium stiffness and LVEDP. We have developed a pipeline to construct a finite-element (FE) representation of a human LV from magnetic resonance imaging (MRI). An inverse modeling approach was used to estimate the myocardial stiffness and LVEDP from regional cardiac strains. Our results indicated that such an approach can closely approximate these important values using cardiac kinematic behavior that is available from established imaging protocols. This initial work points towards the possibility of the accurate estimation of stiffness and pressure from image-derived metrics.

\begin{figure}[htbp]
\centerline{\includegraphics[width=3.5in]{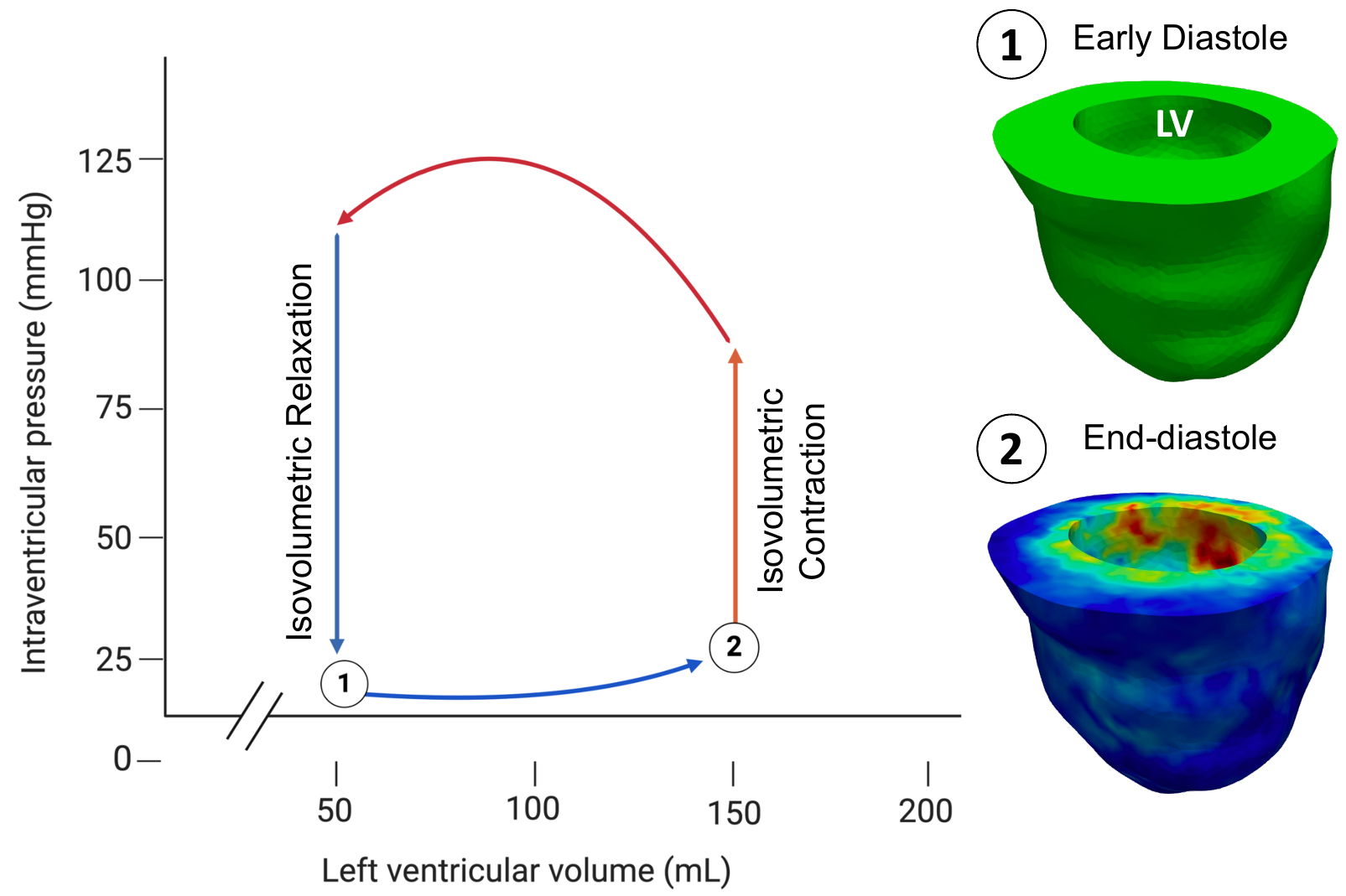}}
\caption{The inverse problem approach involves (1) the reconstruction of the left ventricular geometry at early diastole from cine imaging, followed by the assignment of myocardial stiffness and end-diastolic pressure (EDP). Next, (2) loading the geometry at early diastole with intraventricular pressure iteratively. The cardiac strains are calculated and compared to the ``ground-truth'' strains obtained apriori using EDP. The process is iterated to minimize the error between the simulated ED strains with the ground-truth strains, yielding optimized values for the stiffness parameter and EDP.}
\label{fig:PVloop}
\end{figure}

\section{Methodology}

\subsection{Image reconstruction}
High-resolution cardiac cine magnetic resonance imaging (cMRI) for a LVDD human patient was used for creating the LV geometry. Scans were performed using a Siemens Magetom Skyra Fit scanner at an isotropic resolution of 1.5 mm. Segmentation and reconstruction of the isolated LV at the end of isovolumetric relaxation (Fig. \ref{fig:PVloop}) were performed using Mimics Innovation Suite (Materialise, Leuven, Belgium). The 3-D LV geometry was then truncated below the valve plane. The final LV geometry was meshed using linear tetrahedral elements.

\subsection{Material constitutive model}
The myocardium was modeled as a hyperelastic, transversely anisotropic, and nearly incompressible material. The constitutive model takes the form of the following modified exponential Fung model \cite{Guccione1991, Avaz2019}

\begin{equation} \label{W}
W({\bf E})=\frac{c}{2}\,\left\{e^ {Q} -1 \right\},
\end{equation}

where $Q$ is expressed as

\begin{equation}
    Q = \left[B_1\,{E}_{11}^2 + B_2\,({E}_{22}^2 + {E}_{33}^2 + {E}_{23}^2) + B_3\,({E}_{12}^2 + {E}_{13}^2)\right].
\end{equation}

In these equations, ${\bf E}$ is the Green-Lagrange strain tensor, and $c$, $B_1$, $B_2$, and $B_3$ are material constants. The strain components refer to the local preferred material directions with the axis "1" denoting the fiber direction. The values of the $B_i$ constants governing the local anisotropic behavior of the myocardium were taken from a previous study \cite{Nikou2015}.

In the absence of subject-specific architectural data, a synthetic fiber architecture was applied to the LV model with the fiber angle varying transmurally from $-60^o$ at the epicardium to $60^o$ at the endocardium \cite{Geerts2002}.

\begin{figure}[htbp]
\centerline{\includegraphics[width=3.5in]{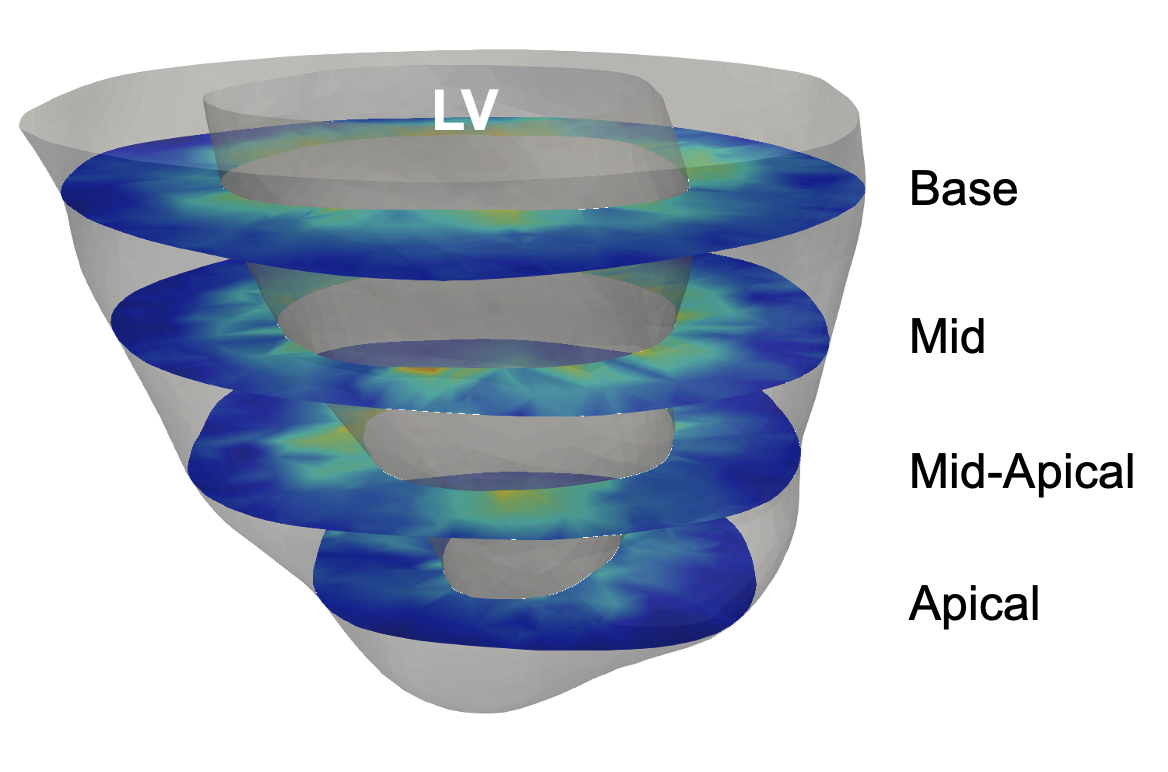}}
\caption{Anatomical strains were calculated and averaged on four short-axis cross-sections of the LV model.}
\label{fig:SA_slices}
\end{figure}

\subsection{Synthetic ground-truth data}
A forward problem was performed to generate synthetic ``ground-truth'' strain data, which was used in the subsequent inverse problem. The stiffness parameter $c$=296.0 Pa was used in the forward problem, and the FE model was loaded to the $EDP$=9.0 mmHg (Fig. \ref{fig:PVloop}). Mean anatomical strains were calculated at ED on four short-axis cross-sections of the LV (Fig. \ref{fig:SA_slices}). These calculated strains were used as the target (ground-truth) strains in the inverse problem described below.

\subsection{Inverse problem}
An inverse modeling approach was used to estimate the LV myocardium stiffness ($c$) and $EDP$ by matching the previously calculated ground-truth strains. Mean circumferential, radial, and longitudinal strains values in the endocardial regions were calculated for each individual cross-section. Endocardial strains were chosen due to their higher sensitivity to the myocardial stiffness in diastole (Fig. \ref{fig:SA_slices}). A Livenburg-Marquet least square minimization scheme was used to iteratively minimize the sum of the squared differences between the respective ground-truth and simulated strain values following a threshold error< 0.01\%. Parameters were constrained to 0 - 1000 Pa and 0 - 25 mmHg for the stiffness parameter $c$ and $EDP$, respectively. 

\begin{figure}[htbp]
\centerline{\includegraphics[width=3.5in]{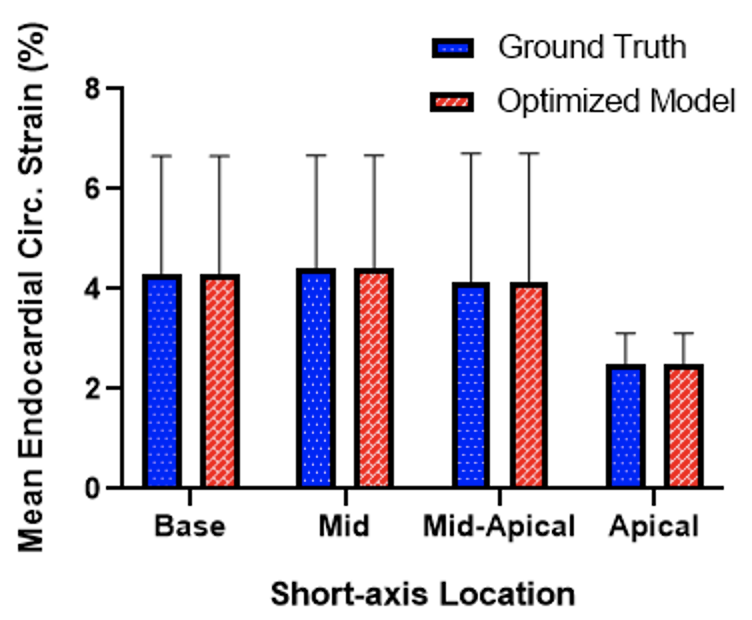}}
\caption{The optimized model closely reproduced the mean circumferential strain at each short-axis location. Multiple t-test analysis indicated no significant difference between the values from ground-truth and the optimized model.}
\label{fig:mean_strains}
\end{figure}

\section{Results}

\subsection{Ventricle kinematics}
\begin{figure}[h!]
\centerline{\includegraphics[width = 0.8\columnwidth]{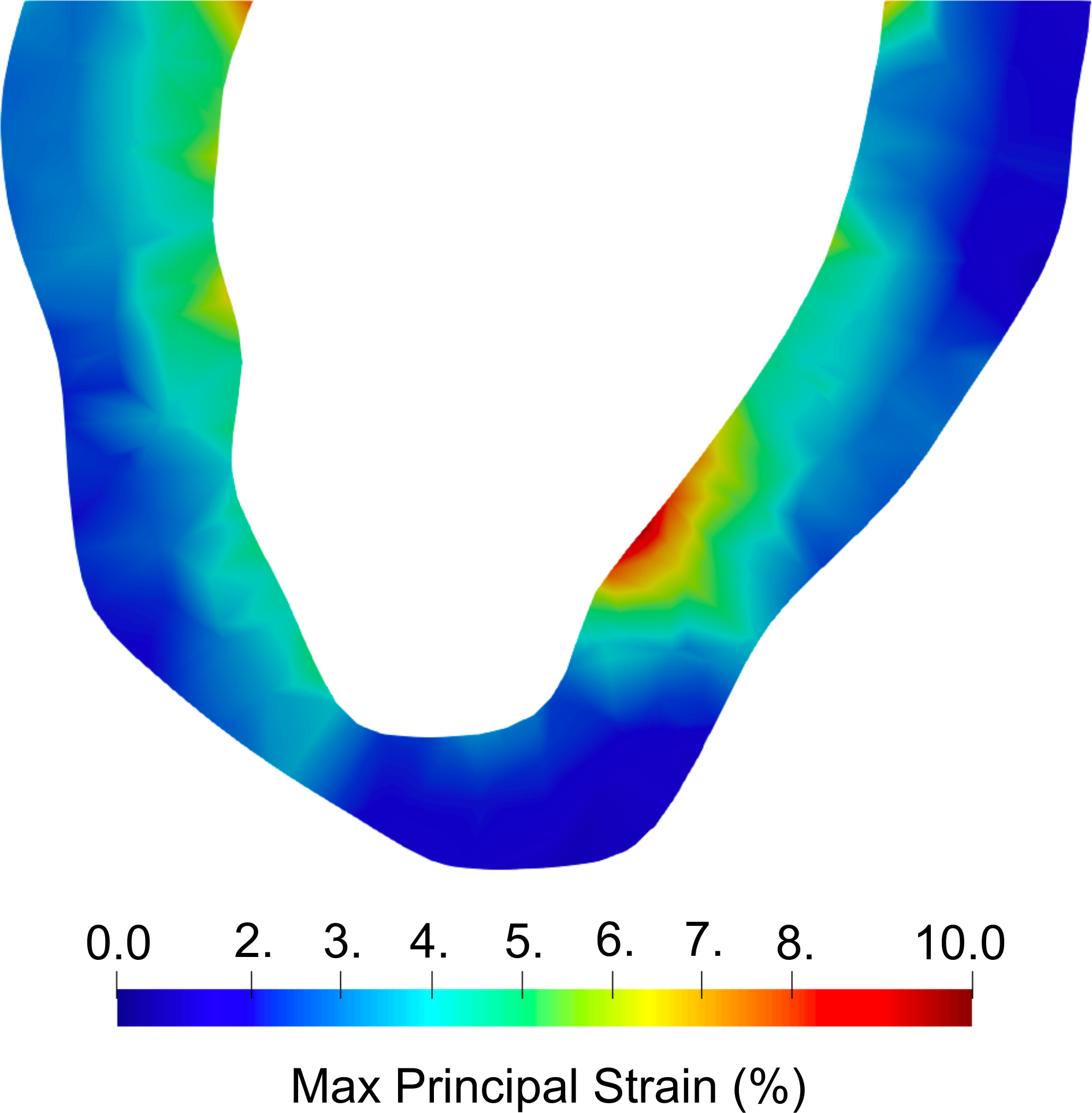}}
\caption{Visualization of the max principal strain at end-diastole on a longitudinal cross-section of the left ventricle model.}
\label{fig:principal_strain}
\end{figure}

The ground-truth LV model indicated mean endocardial circumferential strains of 4.3\%, 4.4\%, 4.1\%, and 2.5\% at the base, mid, mid-apical, and apex SA slices, respectively (Fig. \ref{fig:mean_strains}). The regional distribution of strain indicated higher strain at the endocardium and a transition towards lower strain at the epicardium, consistent with previous experimental studies \cite{Howard2016, Palit2015}. Similarly, the maximum principal strain peaked at $\sim$10\% at the endocardium and transitioned to lower values at the epicardium (Fig. \ref{fig:principal_strain}).

\subsection{Optimized parameters}
\begin{figure}[h!]
\centerline{\includegraphics[width = 0.9\columnwidth]{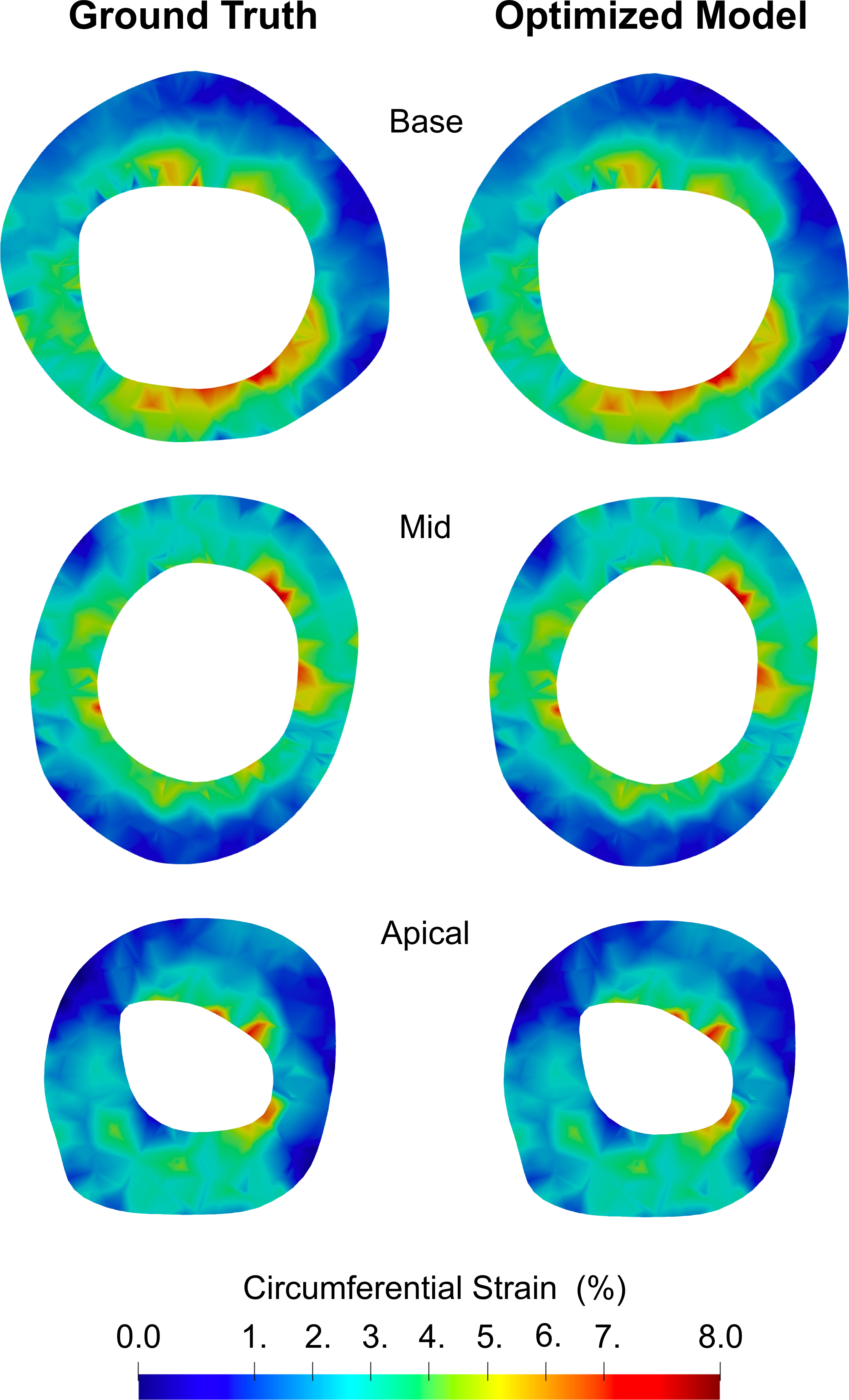}}
\caption{Visualization of the circumferential strain distribution on the base, mid, and apical short-axis cross-sections at end-diastole. An excellent agreement between ground-truth data and optimized model estimations was found.}
\label{fig:strains}
\end{figure}

The inverse model was able to closely estimate the ground-truth stiffness and $EDP$ values (Table \ref{Tab:Props}). The mean ground-truth endocardial strains for each cross-section were accurately matched by the optimized model (Fig. \ref{fig:mean_strains}), with no significant difference between the mean values of the ground-truth and optimized model. Despite only using mean strain values in the inverse problem, the optimized model closely approximated the regional strain profiles of the ground-truth strains (Fig. \ref{fig:strains}).

\begin{table}[h]
	\centering
	\caption{Comparison between ground-truth parameters and parameters estimated via inverse modeling.}
	\begin{tabular}{ccc}
		\toprule[1.5pt]
		   & Ground Truth & Estimated\\
        \hline
        \rule{0pt}{3ex} $c$ (Pa) & 296.0 & 290.7 \\
        $EDP$ (mmHg) & 9.0 & 9.2\\
		\bottomrule[1.5pt]
		
	\end{tabular}
	\label{Tab:Props}
\end{table}

\setlength{\belowcaptionskip}{-100pt}

\section{Discussion}
We have presented a modeling platform that incorporates patient-specific imaging for the estimation of myocardial stiffness and LVEDP. To our knowledge, this work is the first to show that intraventricular pressure can be accurately predicted from cardiac strains using in-silico modeling. In a clinical setting, our modeling framework holds promise to improve feasible, non-invasive, and longitudinal assessment of LVDD.

\subsection{Non-invasive pressure estimation}
Elevated LVEDP serves as a key indicator of diastolic dysfunction, reflecting reduced ventricular compliance and/or impaired relaxation \cite{Lalande2008, Tanmay2023LV}. Clinically, assessing LVEDP is crucial for diagnosing and grading the severity of LVDD. However, LV catheterization to measure LVEDP poses discomfort and incurs risk. Additionally, while global measures of LV stiffness exist (such as dP/dV), they similarly must be derived from invasive measurements. In this work, we have shown that image-based patient-specific modeling can be used as a non-invasive alternative to estimate myocardium stiffness and LVEDP. Our results indicated it is possible to closely approximate these important values using cardiac strain data readily computed from standard cMRI or, potentially, echocardiography, which is more feasible than cMRI. The predictive accuracy of the modeling platform could be enhanced by the introduction of data from long-axis cross-sections. Additionally, leveraging the principal strains, which offer more sensitivity to LVEDP and myocardial stiffness, could improve the accuracy of the pipeline further. This work suggests that an accurate, non-invasive model-based prediction platform could surrogate invasively measured LVEDP and LV stiffness for more frequent LVDD monitoring.

\subsection{Strain: composite index of passive function}
The connection between ventricle kinematics, myocardium compliance, and ventricle pressure is central to unraveling the complex mechanics of cardiac function in LVDD. Ventricular strains, representing the deformation of myocardial tissue during the cardiac cycle, serve as crucial indicators of dynamic and regional changes occurring in the heart. Both myocardium stiffness and LV filling pressure, among other biomechanical mechanisms, directly influence these strain patterns. Understanding this interplay is essential for comprehending the current state of myocardial compliance and the longitudinal myocardial adaptation in response to diastolic dysfunction.

In the presence of LVDD, where impaired relaxation and increased myocardial stiffness are prominent, the adaptive biomechanical mechanisms within the myocardium become crucial determinants of LV diastolic function. Anisotropic stiffening of the myocardium, brought on by regional fibrosis or myofiber stiffening, is one of the primary mechanisms of LV restrictive filling. Another, often overlooked, factor modulating the anisotropic myocardial behavior is the potential adaptation of the LV tissue architecture, which would be reflected in strain distribution alterations. As a result of the intrinsic connection between myocardium structure and composition with cardiac deformation, it is possible such events could be evaluated through the use of strain analysis.

Overall, understanding the intricate effects of each adaptive mechanism in the pathological setting of LVDD is pivotal. This nuanced understanding, aided by diastolic strain analysis, not only deepens our knowledge of LVDD but also holds promise for targeted therapeutic interventions aimed at modulating these adaptive mechanisms to ameliorate diastolic dysfunction.

\subsection{Future directions}
Future work will leverage previously developed image registration methods \cite{Maziyar2021} to calculate ground-truth cardiac strains directly from cMRI. Implementation of such methods will ease the translation of this tool to the clinical setting. Similarly, in-silico platforms can be developed to estimate right ventricular pressure non-invasively to facilitate the assessment of pulmonary hypertension \cite{Mendiola2023RV, Tanmay2023, Jang2017}. In addition, machine learning-based surrogates of FE models \cite{Babaei2022, Mehdi2023, Motiwale2023} can be developed to promote the translation of these in-silico technologies to the clinic.

\section{Conclusions}
In this work, we have shown that a metric characterizing myocardial mechanical behavior and LVEDP can be estimated from cardiac kinematics. Our modeling approach was capable of using cardiac strains to estimate both myocardial stiffness and LVEDP. The introduction of such modeling platforms into clinical practice could reduce the need for invasive tests and improve the longitudinal assessment of LVDD.


\section*{Acknowledgment}
R.A. would like to thank the support from the NIH (R00HL138288) and the NSF (2244995).

\section*{Copyright}
$\copyright$ 2024 IEEE.  Personal use of this material is permitted. Permission from IEEE must be obtained for all other uses, in any current or future media, including reprinting/republishing this material for advertising or promotional purposes, creating new collective works, for resale or redistribution to servers or lists, or reuse of any copyrighted component of this work in other works.

\vspace{12pt}

\end{document}